\documentclass[preprint,superscriptaddress,prl]{revtex4-1}
\usepackage{graphicx}
\usepackage{amsmath,amssymb}
\usepackage{mathrsfs}
\usepackage[normalem]{ulem}
\usepackage{color}
\usepackage{soul}
\usepackage{hyperref}
\usepackage[T1]{fontenc}
\usepackage[utf8]{inputenc}
\usepackage{bm}
\usepackage[left]{lineno}

\newcommand{\be}{\begin{equation}}
\newcommand{\ee}{\end{equation}}
\newcommand{\bea}{\begin{eqnarray}}
\newcommand{\eea}{\end{eqnarray}}

\newcommand{\vbg}{V_{{\rm BG}}}
\newcommand{\vsd}{V_{{\rm SD}}}
\newcommand{\isd}{I_{{\rm SD}}}
\newcommand{\mo}{\rm MoS_{{2}}}

\newcommand{\si}{\rm{SiO_{2}}}

\newcommand{\ignore}[1]{}
\newcommand{\UFMG}{Departamento de Fisica, Universidade Federal de Minas Gerais, Belo Horizonte, MG 31270-901, Brasil}
\newcommand{\Cambridge}{Cambridge Graphene Centre, University of Cambridge, Cambridge CB3 0FA, United Kingdom}

\renewcommand{\phi}{\varphi}
\renewcommand{\epsilon}{\varepsilon}

\usepackage{fancyhdr}
\begin{document}

\title{Local photodoping in monolayer MoS2}

\author{Andreij C. Gadelha}
\affiliation{\UFMG}

\author{Alisson R. Cadore}
\affiliation{\UFMG}
\affiliation{\Cambridge}

\author{Lucas Lafeta}
\affiliation{\UFMG}

\author{Ana M. de Paula}
\affiliation{\UFMG}

\author{Leandro M. Malard}
\affiliation{\UFMG}

\author{Rodrigo G. Lacerda}
\affiliation{\UFMG}

\author{Leonardo C. Campos}
\affiliation{\UFMG}

\begin{abstract}
\date{\today}

\textbf{Inducing electrostatic doping in 2D materials by laser exposure (photodoping effect) is an exciting route to tune optoelectronic phenomena. However, there is a lack of investigation concerning in what respect the action of photodoping in optoelectronic devices is local. Here, we employ scanning photocurrent microscopy (SPCM) techniques to investigate how a permanent photodoping modulates the photocurrent generation in MoS$_{\mathbf{2}}$ transistors locally. We claim that the photodoping fills the electronic states in MoS$_{\mathbf{2}}$ conduction band, preventing the photon-absorption and the photocurrent generation by the MoS$_{\mathbf{2}}$ sheet. Moreover, by comparing the persistent photocurrent (PPC) generation of MoS$_{\mathbf{2}}$ on top of different substrates, we elucidate that the interface between the material used for the gate and the insulator (gate-insulator interface) is essential for the photodoping generation. Our work gives a step forward to the understanding of the photodoping effect in MoS$_{\mathbf{2}}$ transistors and the implementation of such an effect in integrated devices.}
\end{abstract}

\maketitle
\thispagestyle{fancy}
\subsection*{Introduction}
\lhead{}
Two-dimensional (2D) materials are strategically important for optoelectronic applications due to their ultra-thin nature and tunable electrostatic and optical properties \cite{slmt,2dreview,ultra,ultraviolet,slmp,hdm,pnreview,andreij,tvhe,Schneider2018,Lundt2019,monoopt,2dreview,pnreview}. Among all layered materials \cite{Mounet2018}, the vast family of the transition metal dichalcogenides (TMDCs) has particular significance because many of these materials are direct band gap 2D semiconductors in the monolayer form \cite{2dreview}, bringing light to interesting valleytronic properties \citep{Schneider2018,Lundt2019,tvhe}, high photoluminescence emission \cite{Kimeaau4728}, and high photocurrent response \cite{ultra,ultraviolet,slmp,hdm,pnreview}. Furthermore, many paths have been proposed to tune the optoelectronic properties of TMDCs, including strain engineering \cite{strain,Palacios-Berraquero2017,GANT20198}, electric and magnetic fields applications \cite{Li2019,Barbone2018}, and more recently integrating TMDC monolayers in twisted heterostructures \cite{Lu2019,moireex}. In special, using laser exposure to modulate the density of charge in 2D materials, also called by photodoping effect, has emerged as a convenient way to engineer optoelectronic devices \cite{andreij,Epping_2018,phomote,photpe,dinmo,repho,monoopt,pattws2,etpp,fepn,Quereda_2019,Martinez2017,nonpropho,memopn}. For instance, photodoping has been used to generate p-n junctions in graphene \cite{fepn}, for applications in photovoltaics using WSe$_{2}$ \cite{memopn}, to enhance the optoelectronic performance of WS$_{2}$ \cite{pattws2}, to demonstrate a gate-tunable photomemory effect \cite{andreij}, and to propose alternative optical memory devices \cite{memopn,monoopt,nonpropho,multiMoop}. However, there is a lack of information regarding how photodoping modifies TMDCs in a locally and controllable fashion.

Here we investigate the local modification of $\mo$ photocurrent due to the photodoping effect in a monolayer $\mo$ on top of SiO$_{2}$/Si transistor. We demonstrate photodoping as high as $\Delta n_{\textrm{ph}}=\mathrm{8.5\times 10^{12}\,cm^{-2}}$ in $\mo$ just by laser exposure and gate-voltage applications. Also, temporal analysis of the photodoping suggests that it is a permanent doping, retaining up to 77~\% of the initial photodoping. Besides, we use a scanning photocurrent microscopy setup (SPCM) that gives information about the photodoping effect with a resolution limited by the spot area of the laser (approximately 1~$\mu\mathrm{m}^{2}$ in our setup). Then, we show that the photodoping is a local effect and inhibits photocurrent in specific local regions of monolayer $\mo$. Finally, we study the persistent photocurrent (PPC) in the $\mo$ sheet placed in different substrates, and we point out that the gate-insulator interface is crucial for the photodoping generation. In summary, our work expands the possibilities of controlling the optoelectronic properties of 2D materials.

\subsection*{Results and Discussion}
\begin{figure*}[!hbtp]
\centering
\centerline{\includegraphics[width=13.6cm]{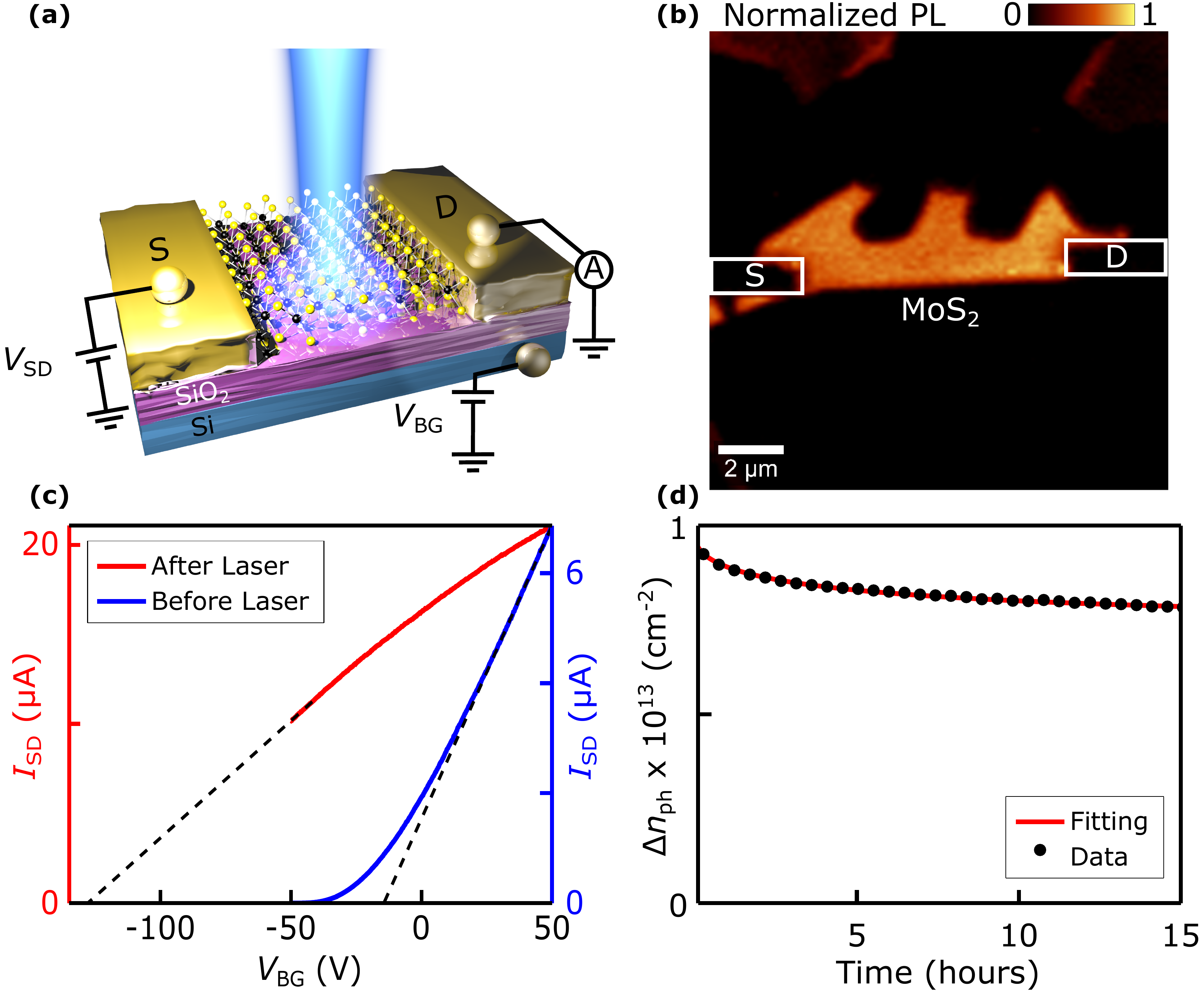}}
\caption{\label{fig:1}\textbf{Photodoping in MoS$_{\mathbf{2}}$}. \textbf{(a)}, sketch of the $\mo$ FET. \textbf{(b)}, normalized photoluminescence image from the integrated area of the intensity of the maximum photoluminescence peak of $\mo$. This measurement is done with a
457 nm laser,  1~$\mu$m spot size and power density of 340~$\mu$W/$\mu$m$^{2}$. \textbf{(c)}, the $\isd$ vs $\vbg$ measurements before (blue) and after (red) the laser exposure, $V_{\mathrm{SD}}=\mathrm{1\,V}$. The red curve is measured after the 488~nm laser exposure with a power density of 700~$\mu$W/$\mu$m$^{2}$ and $\vbg=\mathrm{-50\,V}$ until photocurrent saturation. \textbf{(d)}, photocurrent decay after the photodoping induced by the 488~nm laser with a power density of 700~$\mu$W/$\mu$m$^{2}$ until photocurrent saturation. The parameters $V_{\rm{BG}}=\rm{0\,V}$ and $V_{\rm{SD}}=\rm{1\,V}$ are used for this measurement.}
\end{figure*}

Fig. \ref{fig:1}(a) shows a sketch of the device that we study in this work. The device is a $\mo$ field-effect transistor (FET) composed of a monolayer $\mo$ on a SiO$_{2}$/Si substrate, where we use a highly n-type doped Si wafer as a back gate terminal \cite{natalia}. We do all the measurements in a vacuum (P~$\approx10^{-6}$~mbar) and at room temperature. Fig. \ref{fig:1}(b) shows a map of the intensity of the $\mo$ photoluminescence centered in $\lambda=680\,\rm nm$ and with an excitation wavelength of $\lambda=457\,\rm nm$, demonstrating that the monolayer $\mo$ sheet is uniform. To measure the photodoping, we compare transfer curves, current ($\isd$) vs back gate voltage ($\vbg$) curves, before and after the laser exposure, see Fig. \ref{fig:1}(c). 

 In blue, Fig. \ref{fig:1}(c), we plot a transfer curve before the laser exposure. We estimate the threshold voltage ($V\rm_{th}$) as $V\rm_{th}^{0}=\rm{-14\,V}$ by extrapolating the $\isd$ curve. Then, we expose the $\mo$ device to the laser $\lambda=\mathrm{488\,nm}$ with $\vbg=\mathrm{-50\,V}$ until the photocurrent saturates. Next, we turn the laser off and repeat the measurement to evaluate the transfer curve. The data from the transfer curve after the laser exposure (red curve in Fig. \ref{fig:1}(c)) exhibits a significant increase of $\isd$ for all applied gate conditions. A shift of $V\rm_{th}$ in the experiment causes the increasing behavior of $\isd$, a signature of photodoping effect \cite{andreij,fepn}. Hence, the photodoping effect is a consequence of a change on the density of free charges of $\mo$ promoted by the laser exposure. We estimate that the initial $V\rm_{th}^{\rm{0}}=\rm{-14\,V}$ shifts to $V\rm_{th}^{L}=\rm{-127\,V}$ after laser exposure, see Fig. \ref{fig:1}(c), resulting in a change in the density of charge of $\mo$ of $\Delta n_{\textrm{ph}}=\mathrm{8.5\times 10^{12}\,cm^{-2}}$, which is evaluated using the equation:

\begin{equation} \label{eq:n}
 \Delta n_{\textrm{ph}}=\frac{\epsilon_{0}\epsilon_{ox}}{e\,d}(V\rm_{th}^{L}-V\rm_{th}^{0})
\end{equation} 
where $\epsilon\rm_{ox}$ and $d$ are the dielectric constant of the insulator and its thickness, respectively. Note that such extra doping is obtained simply by the combination of laser exposure and the applied gate bias. 

To exploit the photodoping effect in robust integrated devices, they must hold the photodoping for long periods \cite{nonpropho}. To assess this issue, we measure the $\mo$ photodoping time dependency, see the black dots in Fig. \ref{fig:1}(d). In this case, we evaluate the photodoping by measuring the $\mo$ photocurrent applying $\vsd=\mathrm{1\,V}$ and $\vbg=\mathrm{0\,V}$, and using formula $\Delta n_{\textrm{ph}}= \Delta \sigma_{\textrm{ph}}/e\mu$, where $ \Delta \sigma_{\textrm{ph}}$ is the photoconductance and $\mu$ is the mobility of the device. We measure the decay of the photodoping following the saturation of the photocurrent by laser exposure. After $\mathrm{15\,h}$, the photodoping barely decreases, suggesting that the photodoping effect is permanent. To provide an estimative for the photodoping loss over a long period, we employ an exponential decay fit, the red line in Fig. \ref{fig:1}(d). From the fitting, we predict that the reminiscent photodoping is approximately 77\% of its initial value. Thus, the devices can retain 77\% of the charge induced by the laser permanently, which is ideal for their integration in optoelectronic circuits. 

\begin{figure*}[btp]
\centering
\includegraphics[width=14.4cm]{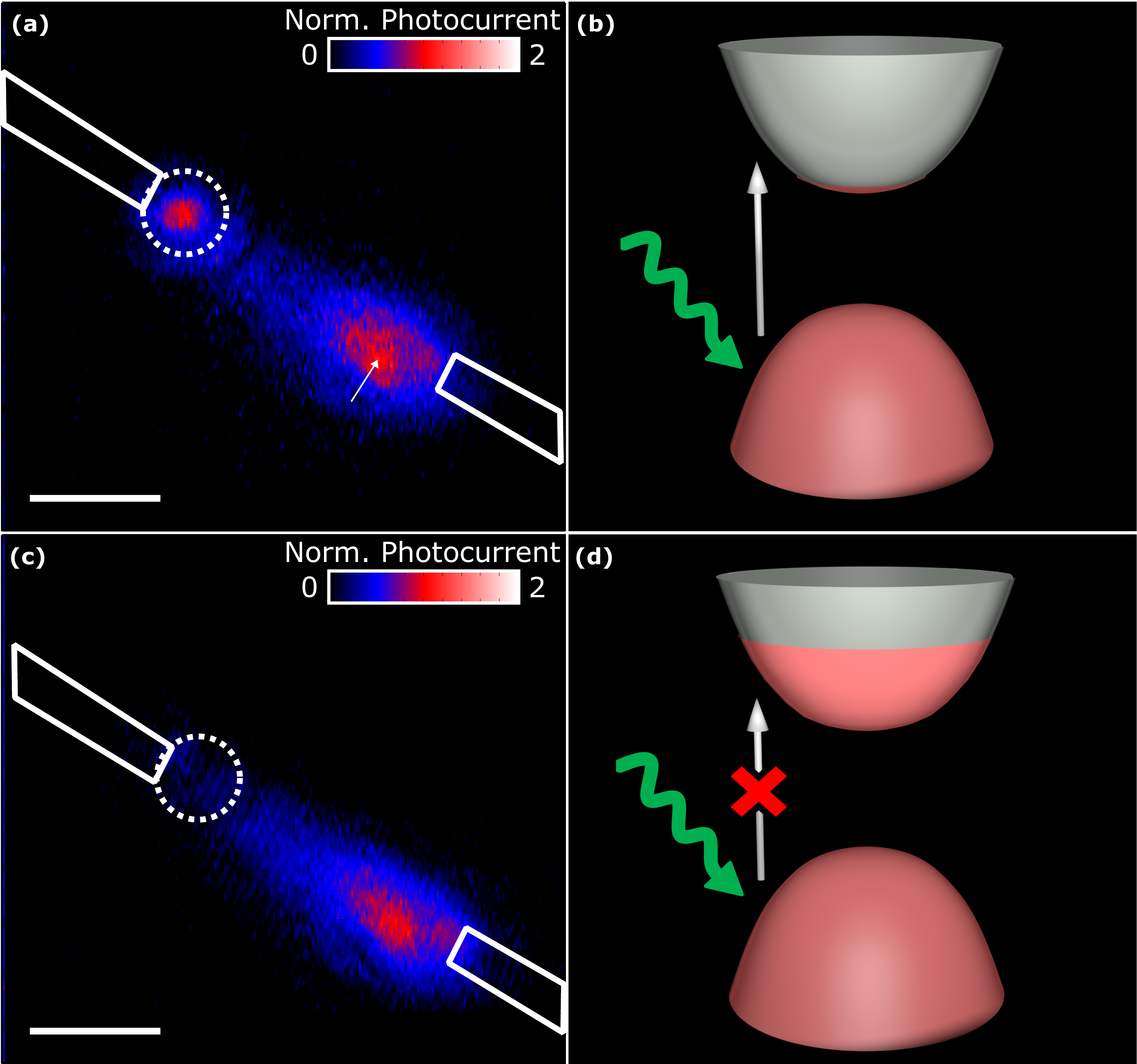}
\caption{\label{fig:3}\textbf{Spatial control of the photocurrent}. \textbf{(a)}, scanning photocurrent microscopy measurement for the $\mo$/SiO$_{2}$ sample. The excitation laser wavelength is 561~nm with a power density of 100~$\mu$W/$\mu$m$^{2}$. In this measurement we use $V_{\mathrm{BG}}=\mathrm{-70\,V}$ and $V_{\mathrm{SD}}=\mathrm{1\,V}$. \textbf{(b)}, the band structure of the region inside the dashed circle in panel \textbf{(a)}. \textbf{(c)}, SPCM measurement from the same region and conditions as in \textbf{(a)}, but after exposing the region inside the dashed circle in panel \textbf{(a)} with the $\lambda=\mathrm{561\,nm}$ laser, power density of 1000~$\mu$W/$\mu$m$^{2}$ and $\vbg=\mathrm{-50\,V}$ for five minutes. \textbf{(d)}, the band structure of the region inside the dashed circle in panel \textbf{(c)}. Scale bar is 4 $\mu$m.}
\end{figure*}

We now show that the photodoping is a local effect by the spatial control of the photocurrent generation in $\mo$, see Fig. \ref{fig:3}. To visualize the spatial distribution of photodoping, we use an SPCM setup. SPCM is a useful tool to investigate local optoelectronic effects and have been used to elucidate the photocurrent in $\mo$ \cite{epum}, to observe photothermoelectric effect in both graphene and $\mo$ \cite{thermomos2,thermographene} and to image a p-n junction in graphene \cite{fepn}.  \noindent The photodoping generation is a relatively slow process because the laser induces typical dopings of $\Delta n_{\textrm{ph}}\approx\mathrm{10^{12}\,cm^{-2}}$ after an exposure time of 10~s in a single point \cite{andreij}. On the other hand, the SPCM measures photocurrents with fast response time ($\approx$~ms) solely.  We support this statement on the SPCM setup, where the laser intensity is modulated at 3~kHz with an optical chopper, generating an alternating photocurrent signal in $\mo$. Subsequently, a lock-in, synchronized with the chopper, filters this alternating signal, preventing the observation of photocurrents with a response time higher than $\approx$~ms. Hence, the SPCM does not measure the photodoping, which is a slow process, directly. What we observe in an SPCM image is predominantly the photocurrent due to the generation of electron-hole pairs in $\mo$, which have a typical photocurrent response time of $\approx$~ms \cite{epum,larpho}. However, we show that we can distinguish photodoping in SPCM images indirectly, see Fig. \ref{fig:3}. Fig. \ref{fig:3}(a) shows a SPCM measurement for the $\mo$ device using a $\mathrm{561\,nm}$ (2.2~eV) laser with a power density of $\mathrm{100\,\mu W}/\mu\mathrm{m^{2}}$. A gate voltage of $\mathrm{-70\,V}$ is applied in $\mo$ while the measurement is performed to ensure that the sample is almost depleted of charges. 

During the SPCM measurement, the laser dwells at each point for small periods ($\tau \approx \mathrm{500\,ms}$), which generates a small ($\Delta n_{\textrm{ph}}\approx\mathrm{10^{11}\,cm^{-2}}$) photodoping in the whole sample. To prevent distortions in the SPCM images due to the small photodoping generation during the SPCM measurement, we normalize all the measurements of Fig. \ref{fig:3} with the photocurrent from the region indicated by the arrow. We show in Fig. \ref{fig:3}(b) the band structure near the band edges for the delineated region by the dashed circle (see Fig. \ref{fig:3}(a)) in order to understand the physical mechanism of the photocurrent generation in Fig. \ref{fig:3}(a). Since the sample is almost depleted with charges, the sample absorbs some photons that excite electrons from the valence to the conduction band generating an excess of carriers \cite{slmp,epum}. Afterward, the photogenerated excess of carriers becomes a photocurrent by the applied bias \cite{slmp,epum}. Indeed, the photocurrent is generated throughout the sample, see Fig. \ref{fig:3}(a), but higher values are attained close to the contact interface due to Schottky electric field \cite{epum}.

After measuring the photocurrent spatially, we expose the delimited area (dashed circle) in Fig. \ref{fig:3}(a) with the 561~nm laser with a power density of 1000~$\mu$W/$\mu$m$^{2}$ and $\vbg=\mathrm{-50\,V}$ for five minutes to ensure that a photodoping is generated. After that, we repeat the SPCM measurement with the same parameters used in Fig. \ref{fig:3}(a), and suppression of the photocurrent occurs in the exposed region, as shown in Fig. \ref{fig:3}(c). We explain this local effect qualitatively via a photodoping process. Fig. \ref{fig:3}(d) depicts the band diagram of $\mo$ in the dashed circle area that shows why the photocurrent is suppressed. After inducing the photodoping in the dashed area, the states in the conduction band are filled, which limits the photoexcitation of electron-hole pairs \cite{ultra}. Thus, there is a suppression of the photocurrent in the exposed area (see Fig. \ref{fig:3}(c)), indicating that the photodoping effect controls locally the photocurrent in the $\mo$ channel. 

\begin{figure*}[!hbtp]
\centering
\includegraphics[width=7.2cm]{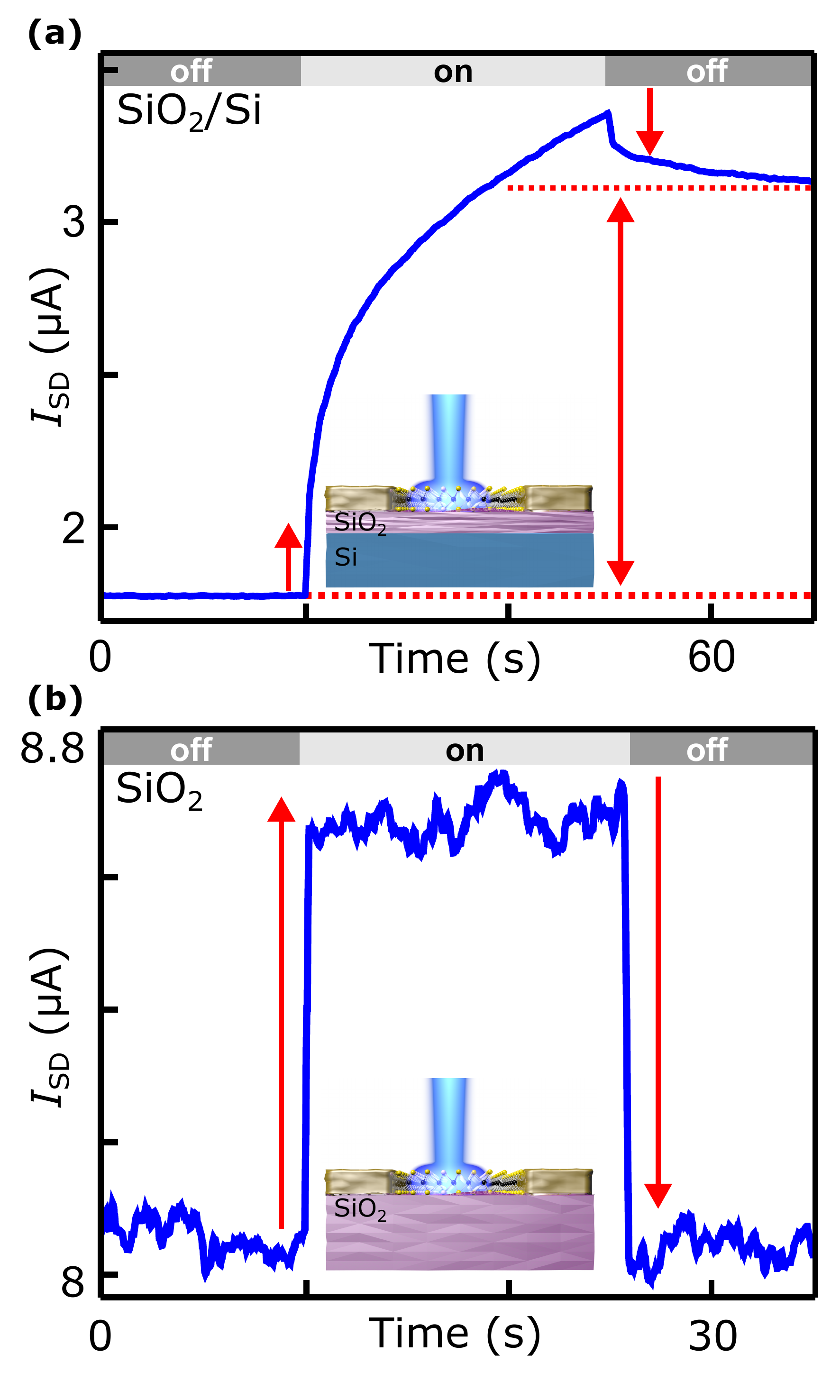}
\caption{\label{fig:2}\textbf{Photodoping and the gate-insulator interface}. \textbf{(a)}, time-resolved photocurrent for the $\si$/Si device, $\vbg\mathrm{=0\,V}$. In the inset, sketch of the $\si$/Si device. \textbf{(b)}, time-resolved photocurrent for the $\si$ device. In the inset, sketch of the $\si$ device. In both measurements we use a 488~nm laser with a power density of 60~$\mu$W$/\mu$m$^{2}$.}
\end{figure*}

Now, we show that interactions with the gate-insulator interface impact the photodoping generation process. We propose in reference \cite{andreij} a model that claims the physics of the photodoping effect is contained mostly in the gate-insulator interface. As the gate-insulator interface is crucial for the photodoping generation \cite{andreij}, we predict that devices that do not have a gate-insulator interface do not generate photodoping. To evaluate this point experimentally, we compare the photodoping generation of two devices, with and without a gate terminal, keeping the same insulator as the substrate, see Fig. \ref{fig:2}. \noindent The simplest way to prepare these devices is by using a standard $\mo$ FET on a SiO$_{2}$/Si substrate (SiO$_{2}$/Si device), and by using a $\mo$ sample on top of glass ($\si$ device). So, the first device has a gate electrode, whereas the second one does not, while both have the same insulator substrate ($\si$). Fig. \ref{fig:2}(a) shows a $\isd$ vs Time curve of the $\si$/Si device, and in the inset we show a sketch of this device. In Fig. \ref{fig:2}(a), we initially measure the current of the device in dark conditions, then we turn the laser on ($\lambda=\mathrm{488\,nm}$) for approximately $\mathrm{25\,s}$.

By analyzing the time dependence of current (Fig. 3(a)), we point out two optical processes that are generating the photocurrent in the $\mo$ channel, see Fig. \ref{fig:2}(a). First, we observe a rapid enhance of $\isd$ due to excitation of electron-hole pairs (see vertical red arrow), followed by a second and slow process that keeps increasing the photocurrent with time. A similar trend, but with decreasing photocurrent occurs when the laser is turned off, there is a fast reduction of $\isd$, due to the recombination of electron-hole pairs, succeed by a slow decay process that leads to a PPC. In reference \cite{andreij}, we argued that the photodoping effect causes the PPC, and here we show experimentally that the gate-insulator interface is essential for the photodoping generation. To elaborate on this issue, we analyze the photocurrent generation in the $\si$ device. Fig. \ref{fig:2}(b) shows an $\isd$ vs Time curve of the $\si$ device, while the inset shows a sketch of this device. In contrast to the response of the SiO$_{2}$/Si device, the $\si$ device presents a very sharp and fast response. Thus, the $\si$ device, which does not have a gate-insulator interface, presents negligible PPC. So, the choice of the substrate interface in a $\mo$ device is decisive to the observation of a PPC, as reported in previous experiments \cite{extrinsic}. 

\subsection*{Conclusion}
In conclusion, we showed that a high permanent photodoping could be attained in $\mo$ transistors only by laser exposure and applied gate voltage. We used an SPCM setup to demonstrate that the photodoping effect controls the photocurrent generation spatially in the $\mo$ transistor. Moreover, we showed that the photodoping mechanism is only present in devices with a gate-insulator interface. Therefore our work opens up possibilities to use the local photodoping in novel optoelectronic devices and to engineer the gate-insulator interface in order to control the photodoping mechanism.

\subsection*{Methods}
\textbf{Device Fabrication}. The $\si$/Si device was obtained from direct exfoliation of $\mo$ on a Si wafer with a 285~nm thick silicon oxide. Metal leads were patterned by electron-beam lithography and subsequent deposition of metals (Au 50~nm). The $si$ device was obtained by depositing the metals firt (Cr 1~nm/Au 50~nm) on a glass substrate. Then $\mo$ flakes were transferred to this structure by dry viscoelastic stamping technique \cite{PDMS}. 

\textbf{Optoelectronic Measurements}. To provide a source-drain bias the external DC source of a standard lock-in amplifier (SR830) was used. While to provide a gate bias a Keithley 2400 were used. The current of the devices was collected by a pre-amplifier and then measured by a multimeter (Keithley 2000). To generate the photocurrent in the $\mo$ FET a 488~nm laser beam was focused in the devices by a 50$\times$ objective lens ($\sim\mathrm{1\,\mu m}$ spotsize). For SPCM measurements, a 561~nm laser beam was directed to a chopper working at 3000~Hz and then focused on the device by a 50$\times$ objective lens. The $\mo$ alternate photocurrent was collected by the pre-amplifier and the signal was measured by the lock-in amplifier

\subsection*{Acknowledgments}
This work was supported by CAPES, Fapemig (Rede 2D), CNPq, Rede de Nano-Instrumentação, and INCT/Nanomaterials de Carbono.
The authors are thankful to the Laboratory of Nano Spectroscopy at UFMG for providing an experimental setup for this work, and to Centro Brasileiro de Pesquisas Fisicas (CBPF) and Centro de Componentes Semicondutores (CCS) for providing an e-beam lithography system and the Lab Nano at UFMG for allowing the use of an atomic force microscope.

\subsection*{Competing financial interests}
The authors declare no competing financial interests



\subsection*{References}
\bibliographystyle{unsrt}

\end{document}